\pdfoutput=1
\documentclass[journal,10pt]{IEEEtran}
\normalsize

\ifCLASSINFOpdf
 \usepackage[pdftex]{graphicx}
\else
 \usepackage[dvips]{graphicx}
\fi

\DeclareGraphicsExtensions{.eps}
\usepackage[cmex10]{amsmath}
\usepackage{bm}
\usepackage{upgreek}
\usepackage{epsfig}
\usepackage[noadjust]{cite}
\usepackage{latexsym}
\usepackage{multirow}
\usepackage{epstopdf}
\usepackage{color}  
\usepackage{amssymb}
\graphicspath{{./Figures/}}
\usepackage{color}
\usepackage{bbm}
\usepackage{mathtools}
\usepackage{url}

\usepackage{cite}
\usepackage[tight,footnotesize]{subfigure}
\usepackage{balance}

\usepackage{color, soul}

\usepackage{siunitx}
\DeclareSIUnit\bps{bps}
\DeclareSIUnit\Torr{Torr}
\DeclareSIUnit\torr{Torr}
\DeclareSIUnit\sample{Sa}

\usepackage{cite}

\begin{document}
% TO COUNT LINES PER PAGE
%\setpagewiselinenumbers
%\pagewiselinenumbers 

\title{Combating the Distance Problem in the \\ Millimeter Wave and Terahertz Frequency Bands}

\author{Ian~F.~Akyildiz,~\IEEEmembership{Fellow,~IEEE,}
        Chong~Han,~\IEEEmembership{Member,~IEEE,}
        and~Shuai~Nie,~\IEEEmembership{Student~Member,~IEEE}% <-this % stops a space
\thanks{This work was supported by the U.S. National Science Foundation (NSF) under Grant No. ECCS-1608579, the European Union via the Horizon 2020: Future Emerging Topics call (FET Open) through the Project
VISORSURF under Grant EU736876, and in part by Alexander von Humboldt Foundation through Dr. Ian Akyildiz's Humboldt Research Prize in Germany. This work was also supported by National Natural Science Foundation of China (NSFC) under Grant No. 61701300 and Shanghai Sailing (YANG FAN) Program under Grant No. 17YF1409900. 
% This work was also supported by National Natural Science Foundation of China (NSFC) under Grant No. 61701300.

Ian F. Akyildiz and Shuai Nie are with the Broadband Wireless Networking Lab at School
of Electrical and Computer Engineering, Georgia Institute of Technology, Atlanta,
GA, 30332 USA. E-mail: \{ian, shuainie\}@ece.gatech.edu.}% <-this % stops a space

\thanks{Chong Han is with University of Michigan -- Shanghai Jiao Tong University Joint Institute, Shanghai Jiao Tong University, Shanghai, China, 200240. E-mail: chong.han@sjtu.edu.cn.}}% <-this % stops a space


\markboth{IEEE Communications Magazine, 2018}
{}
%{Shell \MakeLowercase{\textit{et al.}}: Bare Demo of IEEEtran.cls for IEEE Communications Society Journals}



\maketitle

\begin{abstract}
In the millimeter wave  (30-300~GHz) and Terahertz (0.1-10~THz) frequency bands, high spreading loss and molecular absorption often limit the signal transmission distance and coverage range. In this paper, four directions to tackle the crucial problem of distance limitation are investigated, namely, a physical layer distance-aware design, ultra-massive MIMO communication, reflectarrays, and intelligent surfaces. Additionally, the potential joint design of these technologies is proposed to combine the benefits and possibly further extend the communication distance. Qualitative analyses and quantitative simulations are provided to illustrate the benefits of the proposed techniques and demonstrate the feasibility of mm-wave and THz band communications up to 100 meters in both line-of-sight and non-line-of-sight areas. 
\end{abstract}


% \fi
\IEEEpeerreviewmaketitle

\section{Introduction}
\IEEEPARstart{W}{ireless} communication systems are experiencing a revolution due to the change in the way today's society creates, shares and consumes information. 
The need to provide ubiquitous wireless connectivity will result in over 11 billion mobile-connected devices by 2020, mainly in part brought by the Internet of Things (IoT) paradigm~\cite{akyildiz20165g}. In parallel to the growth in the number of interconnected devices, there has been an increasing demand for higher data rates, of at least 100 times beyond current networks; lower latency of around 1 millisecond; reduced energy consumption; improved reliability and security; and higher scalability. Following this trend, hundreds of Giga-bit-per-second (100~Gbps) and even Terabit-per-second (Tbps) links are expected to become a reality within the next five years.

New spectral bands as well as advanced physical layer solutions are required to support these demands for future wireless communications.
Specifically, \textit{millimeter-wave (mm-wave) communication systems (30-300~GHz)} have been officially adopted in 5G cellular systems. 
Several mm-wave sub-bands have been allocated for licensed communications, including 27.5-29.5 GHz, 36-40~GHz, 57-64~GHz, 71-76 GHz and 81-86 GHz, and 92-95 GHz~\cite{ghosh2014millimeter}. 
While the trend for higher carrier frequencies is clear, the total consecutive available bandwidth for mm-wave systems is still less than 10 GHz, which makes it difficult to support Tbps data rates.
In this context, \textit{Terahertz (THz) band (0.1-10~THz) communication} is envisioned as a key wireless technology to satisfy the future demands within 5G and beyond ~\cite{kurner2014towards,akyildizPHYCOM}. For many decades, the lack of THz transceivers and antennas made the THz band one of the least explored frequency ranges in the electromagnetic (EM) spectrum. However, major progress in the last ten years is enabling practical THz communication systems~\cite{jornet2014graphene,graphene}. 

The use of the mm-wave and THz frequency bands will address the spectrum scarcity and capacity limitations of current wireless systems, and enable new applications, including ultra-high-speed indoor wireless links (e.g., for virtual and augmented reality), and wireless backhaul and access in small cell networks~\cite{akyildizPHYCOM}. In addition to macro/micro-scale applications, the THz band will also enable wireless communication among nanoscale machines or nanomachines.
The state of the art in nanoscale transceivers and antennas points to the THz band as their frequency range of operation.
Applications that are enabled by the nanomachines range from advanced health monitoring systems to chemical attack prevention systems, wireless networks-on-chip and the Internet of Nano-Things.

\begin{figure*}[h]
\centering
        \includegraphics[width=0.8\textwidth]{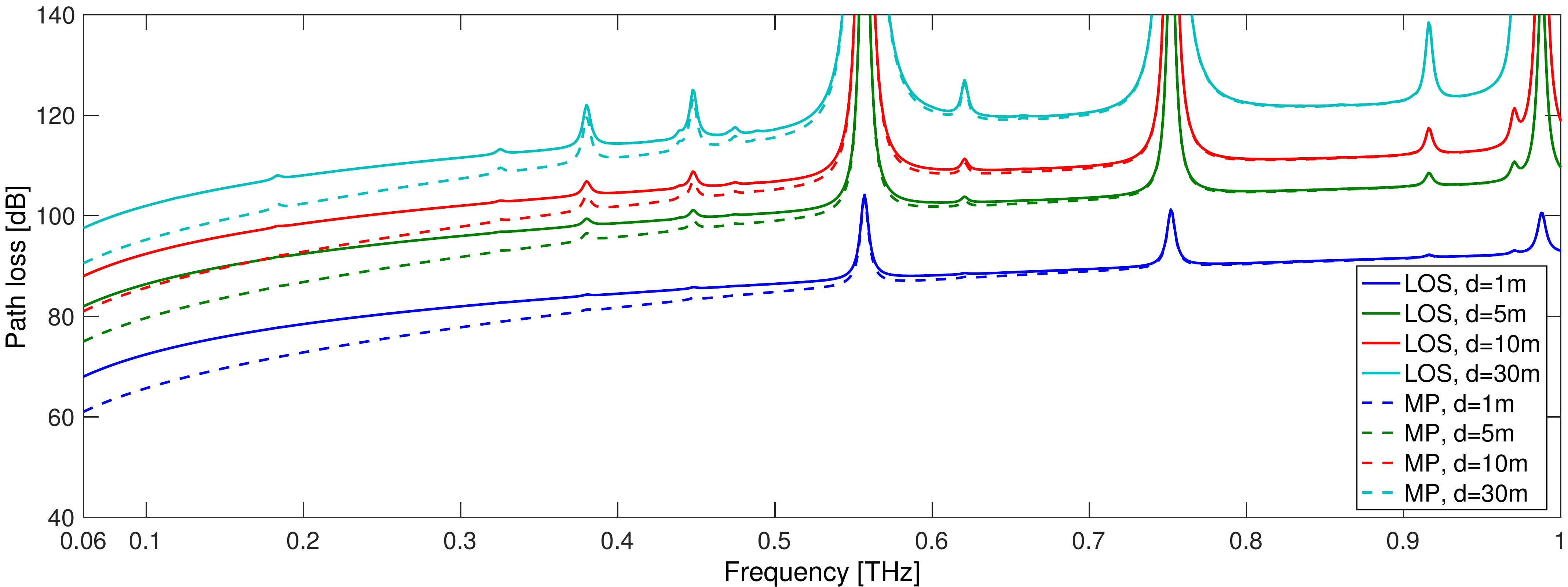} 
         \caption{Path loss of the LoS and multipath channels in the THz band.}
         \label{fig:sw_los}
 \end{figure*}

Nevertheless, a major challenge at mm-wave and THz-band frequencies is posed by the very high propagation loss, which drastically limits the communication distance.
Specifically, the free space path loss of a 10-meter link at THz frequencies can easily exceed 100 dB, which makes communication above tens of meters extremely challenging~\cite{han2016distance}.
On the one hand, the very high path loss arises from the spreading loss, which increases quadratically with the frequency, $\propto f^2$, as defined by the Friis' law.
On the other hand, the molecular absorption loss, which accounts for the attenuation resulting from the fact that part of the wave energy is converted into internal kinetic energy of the molecules in the propagation medium, also contributes to the path loss in the mm-wave and THz bands~\cite{jornet2011channel}. 
Caused by oxygen and water vapor at the mm-wave and THz frequency bands, respectively, the absorption peaks create spectral windows, which have different bandwidths and drastically change with the variation of the distance, as shown in Fig.~\ref{fig:sw_los}. Moreover, despite major developments, the transmission power of THz transceivers is still in the sub-milliWatt range, which further limits the transmission distance.

%
%\begin{figure*}[!h]
%\centering
%        \includegraphics[width=0.5\textwidth]{overview.png} 
%         \caption{An overview of four different directions to combat the distance problem in mm-wave and THz band.}
%         \label{fig:overview}
% \end{figure*}

\begin{table*}[!t]
\centering
\caption{An overview of four different directions to combat the distance problem in mm-wave and THz bands.}
\label{tab:overview}
\begin{tabular}{|c|c|c|c|c|}
\hline
                                                             & \textbf{\begin{tabular}[c]{@{}c@{}}Distance-\\ adaptive design\end{tabular}} & \textbf{UM-MIMO}                                                                                                                                                    & \textbf{Reflectarrays}                                        & \textbf{HyperSurfaces}                                   \\ \hline
\begin{tabular}[c]{@{}c@{}}Theoretical \\ Basis\end{tabular} & Waveform design                                                              & \begin{tabular}[c]{@{}c@{}}Multiple-input\\ multiple-output\end{tabular}                                                                                                      & Ray optics                        & Sub-wavelength ray optics                                                                \\ \hline
Material                                                     & N/A                                                                          & Graphene                                                                                                                                                                 & Patch antenna arrays                               & Metamaterial                                                  \\ \hline
Features                                                     & \begin{tabular}[c]{@{}c@{}}High SNR and \\ distance improvement\end{tabular} & \begin{tabular}[c]{@{}c@{}}High spectral efficiency \\ and distance improvement\end{tabular}  & \begin{tabular}[c]{@{}c@{}}Minimum signal \\ processing\end{tabular}              & \begin{tabular}[c]{@{}c@{}}High spatial resolution \\ and distance improvement\end{tabular}               \\ \hline
Challenges                                                   & Efficiency in dynamic channels                                                                          & \begin{tabular}[c]{@{}c@{}}Implementation \\ complexity\end{tabular}                                                                                                          & \begin{tabular}[c]{@{}c@{}}Limited efficiency \\ for frequency \textgreater 120 GHz\end{tabular}        & High price   \\ \hline
\end{tabular}
\end{table*}

% \subsection{Contributions and Organization of This Paper}

%\subsection{Organization of the paper}
In this paper, we investigate four strategies to combat the distance problem at mm-wave and THz-band frequencies (see Table~\ref{tab:overview}).
Specifically, in Sec.~\ref{sec:modulation}, we discuss the state of the art in terms of \textit{distance-aware physical layer design}, in which the physical layer parameters including the THz spectrum allocation, the pulse waveform, and the transmit power are dynamically adapted according to the transmission distance, and investigate the achievable distance gains~\cite{han2016distance}.
In Sec.~\ref{sec:um-mimo}, we design and analyze the performance of \textit{Ultra-Massive MIMO (UM-MIMO) communications}, which are enabled by the integration of a very large number of antenna elements in very small footprints and are able to increase the communication distance by simultaneously focusing the transmitted signal in space (beamforming) and frequency (absorption-less windows)~\cite{akyildiz2016ummimo}. 
Third, in Sec.~\ref{sec:reflector}, we demonstrate the deployment of \textit{reflectarrays} in the propagation medium which can efficiently overcome the case with line-of-sight (LoS) blockage, and enhance the signal coverage~\cite{hum2014reconfigurable}. 
Fourth, in Sec.~\ref{sec:surface}, we introduce a new type of intelligent surface called \textit{HyperSurfaces} which utilize novel metamaterials with high spatial resolution to extend the transmission distance through the dynamic control of wireless communication environments~\cite{liaskos2015design}.

Furthermore, in Sec.~\ref{sec:joint} we investigate the benefits of jointly designing of the aforementioned technologies.
Based on these presented technologies, we provide numerical evaluations in Sec.~\ref{sec:evaluation} to demonstrate the communication distances of up to 100~meters. Finally, we conclude the paper in Sec.~\ref{sec:conclusion}.

%The paper is organized as follows. Section~\ref{sec:modulation} describes the distance-adaptive physical layer techniques. Section~\ref{sec:um-mimo} introduces the key characteristics of ultra-massive MIMO. Section~\ref{sec:surface} presents the design and architecture of intelligent surfaces. Section~\ref{sec:reflector} introduces two types of reflectors promising to extend the transmission distance. Section~\ref{sec:evaluation} shows numerical evaluation of the presented techniques. Section~\ref{sec:conclusion} concludes the paper.

%%%%%%%%%%%%%%%%%%%%%%%%%%%%%%%%%%%%%%%%%%%%%
\section{Distance-Adaptive Design}\label{sec:modulation}
%The very strong relationship between the distance and the spectral windows at mm-waves and THz frequencies motivates the design to be distance-adaptive~\cite{han2016distance}. Moreover, each spectral window has an ultra-broad bandwidth ranging from multi-GHz to THz, which allows the multi-wideband transmission by dividing each spectral window into many sub-windows. The detailed solution to the resource allocation scheme in the THz network that employs this strategic spectrum allocation principle is provided in~\cite{han2016distance}.
%
%\subsection{Distance-adaptive Multi-wideband Pulse Waveform}
%To combat the frequency selective fading and increase the received SINR, each information symbol on the sub-window is represented by a sequence of very short pulses, which provides the pulse combing gain. Within one sequence, the positions of the pulses are determined by a pseudo-random time-hopping sequence that is specific to each sub-window. 
%By considering the unique characteristics at the THz frequencies and aiming to improve the signal-to-interference-plus-noise ratio (SINR) or equivalently, the distance,
%we form the multi-wideband waveform model~\cite{han2016distance}.

The very strong relationship between the distance and the spectral windows at mm-wave and THz frequencies motivates the development of distance-adaptive communication techniques~\cite{Han2016wideband}. Moreover, each spectral window has an ultra-broad bandwidth ranging from multi-GHz to THz, which can be divided into narrower but still broadband sub-windows and allow parallel multi-wideband transmissions. This property can be leveraged in two ways, namely, distance-adaptive multi-wideband waveform design and distance-aware bandwidth-adaptive resource allocation~\cite{han2016distance}. 

\subsection{Distance-adaptive Multi-wideband Pulse Waveform }

In~\cite{Han2016wideband}, we proposed for the first time to dynamically adapt the waveforms being transmitted to match the channel available bandwidth defined by molecular absorption. This scheme allows the dynamical variation of the rate and the transmit power on each sub-window.
More specifically, we formulate an optimization framework to derive the waveform characteristics and repetition rate which maximize the communication distance while maintaining the rate and transmit power constraints and by taking into account the inter-symbol and inter-band interferences.
The results show that the communication distance can be effectively enhanced as the transmit power and the number of frames increase, at the cost of the power consumption and the rate decrease.

\subsection{Distance-aware Bandwidth-adaptive Resource Allocation}
Besides improving the individual user achievable data rates, the very large available bandwidth can be leveraged to efficiently allocate multiple users. In this direction, in~\cite{han2016distance}, we developed a resource allocation scheme that takes into account the physical parameters including the mm-wave and THz spectrum allocation, the modulation techniques, and the transmit power.  Since the broad bandwidth in the mm-wave and THz band enables very high data rate transmissions, distance maximization becomes the objective, which is different from the traditional resource allocation problem which either minimizes the energy consumption or maximizes the data rate. 

As the distance decreases, the path loss drops and hence, the received power and the usable bandwidth of a spectral window increase. Based on this relationship, the strategic spectrum utilization principle for one spectral window operates as follows. First, the center sub-windows, i.e., the center spectrum in the window, are allocated to the long-distance links. Then, the side sub-windows, i.e., the side spectrum in the window, are allocated to the short-distance transmissions. The resource allocation model that incorporates the aforementioned spectrum allocation principle is formulated in the mm-wave and THz network, and the detailed solution is provided in~\cite{han2016distance}. 

The proposed solutions can effectively increase the communication distance, at the cost of hardware complexity. Currently, only single-band mm-wave and THz-band have been demonstrated. To experimentally demonstrate the proposed solutions, very high-speed digital-to-analog and analog-to-digital converters, DACs and ADCs, respectively, are needed. The fastest converters demonstrated to date can support up to 100 Giga-samples-per-second, which can support transmission bandwidths of up to 50 GHz, i.e., less than the actual bandwidth of the channel. For this, sub-Nyquist sampling strategies are needed. %A distance-aware bandwidth-adaptive resource allocation scheme in the THz band communication network captures the peculiarities of distance-varying spectral windows and efficiently exploits the THz spectrum. 
\section{Ultra-Massive MIMO}\label{sec:um-mimo}
The concept of Massive MIMO has been introduced in recent years, in which the antenna arrays with tens to hundreds of elements are utilized to increase the spectral efficiency. 
The current development of massive MIMO communication has demonstrated a testbed at 28 GHz with 256 antenna elements by Anokiwave in 2017\footnote{Anokiwave Introduces 5G mmWave Reconfigurable 256-Element Active Antenna Array in October 2017. [Online]: http://www.microwavejournal.com/articles/29235-anokiwave-introduces-5g-mmwave-reconfigurable-256-element-active-antenna-array. 
Going one step further, in~\cite{akyildiz2016ummimo}, we introduced for the first time the concept of \textit{ultra-massive MIMO communications $(1024 \times 1024)$}, by equipping 1024 antenna elements at transmitter side and 1024 at receiver side, respectively.
This could be enabled by novel plasmonic nano-antenna arrays and can drastically increase the communication distance at mm-wave and THz frequencies by simultaneously focusing the transmitted signals in space and in frequency.}  As a result, wireless Tbps links can be established over a communication distance of several tens of meters~\cite{akyildiz2016ummimo}.
With current fabrication processes (based on photolithography, electro beam lithography, etc.), the cost of making one front-end or hundreds of them in the same wafer is effectively the same. In addition to the fabrication of such arrays, there will be challenges in the control of the system, which we identify and explain in the section.

%Nanomaterials such as graphene or metamaterials are utilized to build miniature plasmonic antennas~\cite{jornet2013graphene} and transceivers~\cite{jornet2014graphene,akyildiz2016graphene} for very high frequencies. 
%The resulting UM-MIMO $(1024 \times 1024)$ systems can be utilized both at the transmitter and the receiver, to simultaneously overcome the spreading loss problem, by focusing the transmitted signal in space, and the molecular absorption loss problem, by focusing the spectrum of the transmitted signal in the absorption-free windows. 

%Moreover, by creating two-dimensional antenna arrays instead of one-dimensional arrays, the radiation can be controlled both in the elevation and the azimuth directions, thus enabling Full-Dimension (FD) MIMO~\cite{nam2013full}, which have been proved to be very useful for mm-wave communication systems~\cite{swindlehurst2014millimeter,bjornson2016ten}. 
%When moving to the THz band, antennas become even smaller and many more elements can be embedded in the same footprint. 
\subsection{Plasmonic Nano-antenna Arrays}
\label{sec:system}
Plasmonic materials are metals or metal-like materials which support the propagation of surface plasmon polariton (SPP) waves. SPP waves are confined EM waves that appear at the interface between a metal and a dielectric as a result of global oscillations of the electrical charges. Different plasmonic materials can support SPP waves at different frequencies. In particular, \textit{graphene}, a one-atom-thick carbon-based nanomaterial with unprecedented mechanical, electrical and optical properties, has been experimentally proved to support the propagation of SPP waves at THz-band frequencies~\cite{jornet2014graphene,graphene,akyildiz2016ultra}. Metamaterials can support plasmonic waves at mm-wave frequencies~\cite{jornet2014graphene}. Since SPP waves propagate at a much slower speed than EM waves in free space, the resulting wavelength of SPP waves is much smaller than that of EM waves, hence enable the design of a much smaller plasmonic antenna compared to metallic antennas. The advantage of having much smaller sized antennas is the easiness in integrating large 2D antenna arrays which enables the dynamic control of both azimuth and elevation domains efficiently~\cite{akyildiz2016ummimo}. Nevertheless, having so many elements requires the development of innovative ways to operate the array. One option to reduce the number of control elements is to use an array of sub-arrays architecture, but fully digital architectures can potentially be developed, as discussed in~\cite{akyildizPHYCOM}.% In particular, SPP waves propagate at a much lower speed than EM waves in free space. As a result, the SPP wavelength $\lambda_{spp}$ is much smaller than the free-space wavelength $\lambda$. The ratio
%\begin{equation}
%\gamma=\frac{\lambda}{\lambda_{spp}}
%\label{eq:confinement}
%\end{equation}
%is known as the confinement factor and depends on the plasmonic material and the system frequency. The confinement factor $\gamma$ can be obtained by solving the SPP wave dispersion equation with the boundary conditions imposed by the specific device geometry~\cite{jornet2013graphene}. 

%\begin{figure}
%\centering   
%         \includegraphics[width=0.5\textwidth]{Figures/architecture.pdf} 
%        \caption{UM-MIMO using the array-of-subarrays structure in a 2D grid layout. $d_\mathrm{a}$ is the separation among subarrays that is larger than the half wavelength, while $d_\mathrm{sa}$ is the separation among antenna elements in one subarray that is smaller than the half wavelength.}
%         \label{fig:um_mimo} 
%\end{figure}

%By employing graphene-based antenna arrays, an UM-MIMO antenna array architecture is designed as a uniform planar array provides additional degrees of freedom and is able to accommodate a large number of antennas, by placing the antenna elements in both the azimuth and elevation domains, which was also suggested to enable FD-MIMO for cellular technologies. The resulting UM-MIMO schemes can be utilized both at the transmitter and the receiver. 
%Enabled by the small size of individual nano-antennas, large plasmonic nano-antenna arrays can be assembled. In order to achieve control of the radiation in the azimuth and elevation planes, 2D antenna arrays  need to be defined, similarly to FD-MIMO systems in lower-frequency cellular systems. 

\begin{figure*}
\centering   
         \includegraphics[width=0.7\textwidth]{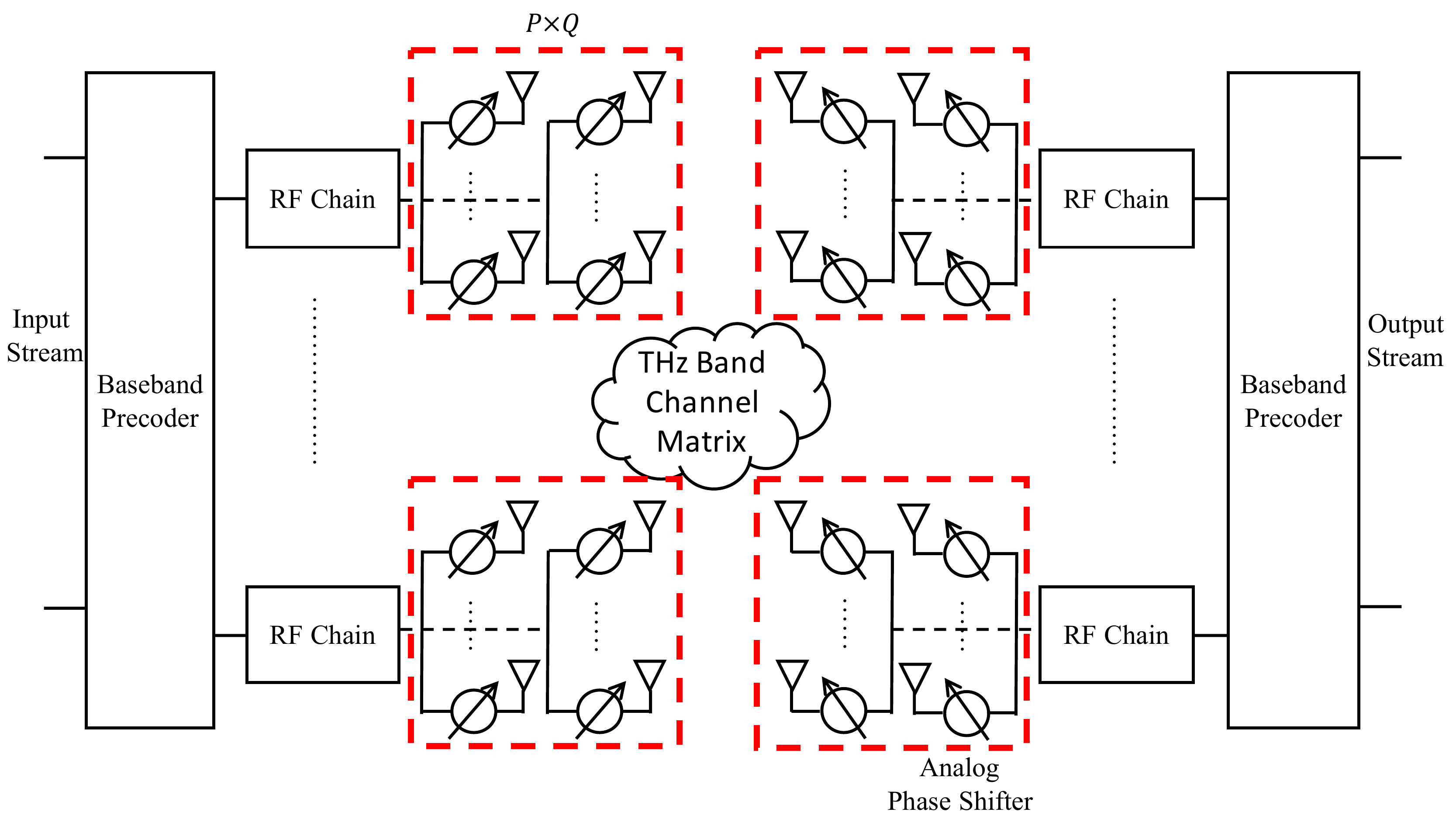} 
        \caption{UM-MIMO communication system architecture with hybrid beamforming. Each subarray is composed of $P\times Q$ tightly packed antenna elements.}
         \label{fig:hybrid} 
\end{figure*}

The resulting very large antenna arrays can be simultaneously utilized in transmission and reception to enable ultra-massive MIMO $(1024 \times 1024)$ schemes.
The conceptual design of an end-to-end UM-MIMO system is illustrated in Fig.~\ref{fig:hybrid}. In particular, $M_t \times N_t$ and $M_r \times N_r$ antenna subarrays are equipped at the transmitter (Tx) and the receiver (Rx), respectively, in which each subarray is driven by an individual baseband-to-RF chain.  Each subarray is composed of $P\times Q$ tightly packed antenna elements. Each antenna element is connected to a wideband THz analog phase shifter, which can be implemented by digitally controlled graphene integrated gates.

%Based on the AoSA architecture, hybrid beamforming can be realized~\cite{lin2015adaptive,alkhateeb2014mimo}, which significantly mitigates the system complexity by reducing the number of RF circuits, while still has the flexibility to provide beamforming and spatial multiplexing gains at the same time. 
%%On the contrary, digital signal processing for precoding and combining allows for advanced transmission strategies, but suffers from high complexity particularly in UM-MIMO systems.
%Moreover, in the AoSA structure, the basic component in the THz system becomes a subarray rather than an antenna element. The beamforming gain enabled in an individual subarray contributes to the overcome the very high path loss at THz frequencies. 

\subsection{Dynamic UM-MIMO Modes}

By properly feeding the antenna array elements, \textit{dynamic array modes} can be adaptively generated. In particular, we investigate different operational modes by grouping the sub-arrays. The large number of antenna arrays can be grouped together and hence is capable of steering high-gain narrow beams towards the strongest propagation path, as beamforming (BF). This beamforming design can effectively overcome the very high attenuation at the mm-wave and THz frequencies and importantly, enhance the communication distance. The spatial multiplexing (SM) scheme uses multiple streams on a single carrier to increase the capacity per user, but is most effective when radio links operate in a high signal-to-noise ratio (SNR) regime and are bandwidth-limited.

The benefits of BF and SM can be achieved at the same time by adopting a combination of these two schemes, thanks to the very large bandwidth in the THz band and the very large antenna arrays with thousands of antenna elements. Particularly, multiple antenna elements can be grouped for BF to increase SNR in power-limited situations. At the same time, unique data streams are provided for the different subarrays on the same carrier frequency for SM to increase the user data rates, as long as the THz channel has sufficiently uncorrelated propagation paths in the spatial domain. 
The tradeoff among the beamforming, the spatial multiplexing or the combination, in terms of the capacity, the spatial degree of freedom and the distance improvement at mm-wave and THz frequencies significantly affects the system performance. The objective is to support multiple ultra-high-speed and long-distance links simultaneously.  

\subsection{Multi-band UM-MIMO}
%Until this point, we have considered a single-band UM-MIMO scheme, which operates at a specific spectral window. However, 
%the huge bandwidth in the THz band can exceed the bandwidth provided by a single antenna~\cite{graphene}. 
To maximize the utilization of the THz channel and enable the targeted Tbps links, more than one spectral window might be needed and utilized at the same time~\cite{akyildizPHYCOM}. Multi-band UM-MIMO enables the simultaneous utilization of different frequency bands. 
More specifically, the response of individual elements in the array can be dynamically and independently modified. 

Following this direction, by using different length antennas in the same array or, in the long term, frequency-tunable graphene-based nano-antennas, the multi-band UM-MIMO systems can be enabled by introducing a new dimension of frequency, which facilitates that multiple transmission windows and even the entire THz band can be simultaneously utilized as a combination of many subbands. 
The advantage is that the multi-band approach allows the information to be processed over a much smaller bandwidth, thereby reducing overall design complexity as well as improving spectral flexibility. 
In this direction, advanced \textit{space-time-frequency (STF) coding and modulation techniques} can be developed for the UM-MIMO systems to exploit all of the spatial, temporal and frequency diversities, and hence, promise to yield remarkable performance improvements.
%
%If the antenna arrays are not available for an individual transmitter, the \textit{cooperative communication} can be exploited to emulate virtual UM-MIMO among a group of transmitters, by enabling multiple sources to simultaneously transmit a common message collaboratively. This technique is specially well-suited for high node density networks (e.g., the Internet of Things). The idea is to control the phase of the transmissions, so that the signals are constructively received at an intended receiver. The transmission range of individual devices can be enhanced at the expense of tight synchronization and coordination requirements among nodes, which remain still as open research challenges.

%%%%%%%%%%%%%%%%%%%%%%%%%%%%%%%%%%%%%%%%%%%%%
\section{Reflectarrays}\label{sec:reflector}
In addition to controlling the radiated signals at the transmitter and the receiver with very large antenna arrays and MIMO communication systems, the utilization of reflectarrays in the propagation medium is also projected to enhance the signal coverage.

Reflectarrays have been widely utilized in radars, point-to-point links, and satellite communications because of their flexibility and low cost~\cite{hum2014reconfigurable}. Based on principles of phased arrays and geometrical optics, electronically tunable reflectarrays can realize dynamically adjustable radiation patterns.  Specifically, the phase shift of of each element in the reflactarray can be controlled electronically and will jointly form an array pattern to receive or transmit the signal to or from desired directions. Compared to phased arrays that require complicated phase shifter circuits and suffer from high transmission line loss at mm-wave and THz frequencies, reflectarrays are simpler in mass production and have higher energy efficiency because there is no need for transmission lines. Since under communication scenarios where receivers are expected to move, antenna arrays should also be flexibly adjustable in order to keep a satisfying SNR at the user end.

\begin{figure*}[!t]
\centering
	\subfigure[]{ \includegraphics[trim = 0mm 0mm 0mm 0mm, clip, width=0.25\textwidth]{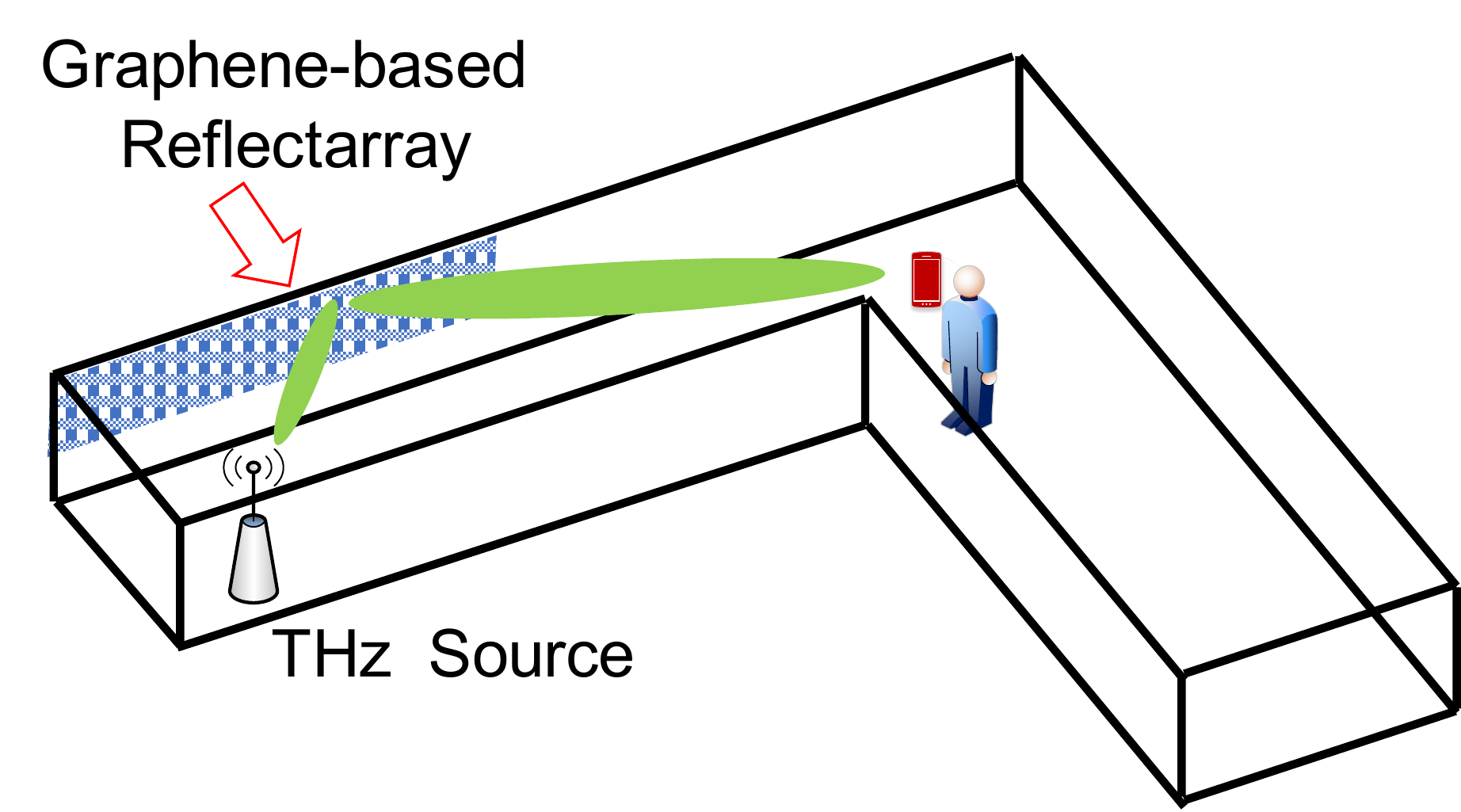} \label{fig:RefArray}}
        \subfigure[]{ \includegraphics[trim = 0mm 0mm 0mm 0mm, clip, width=0.3\textwidth]{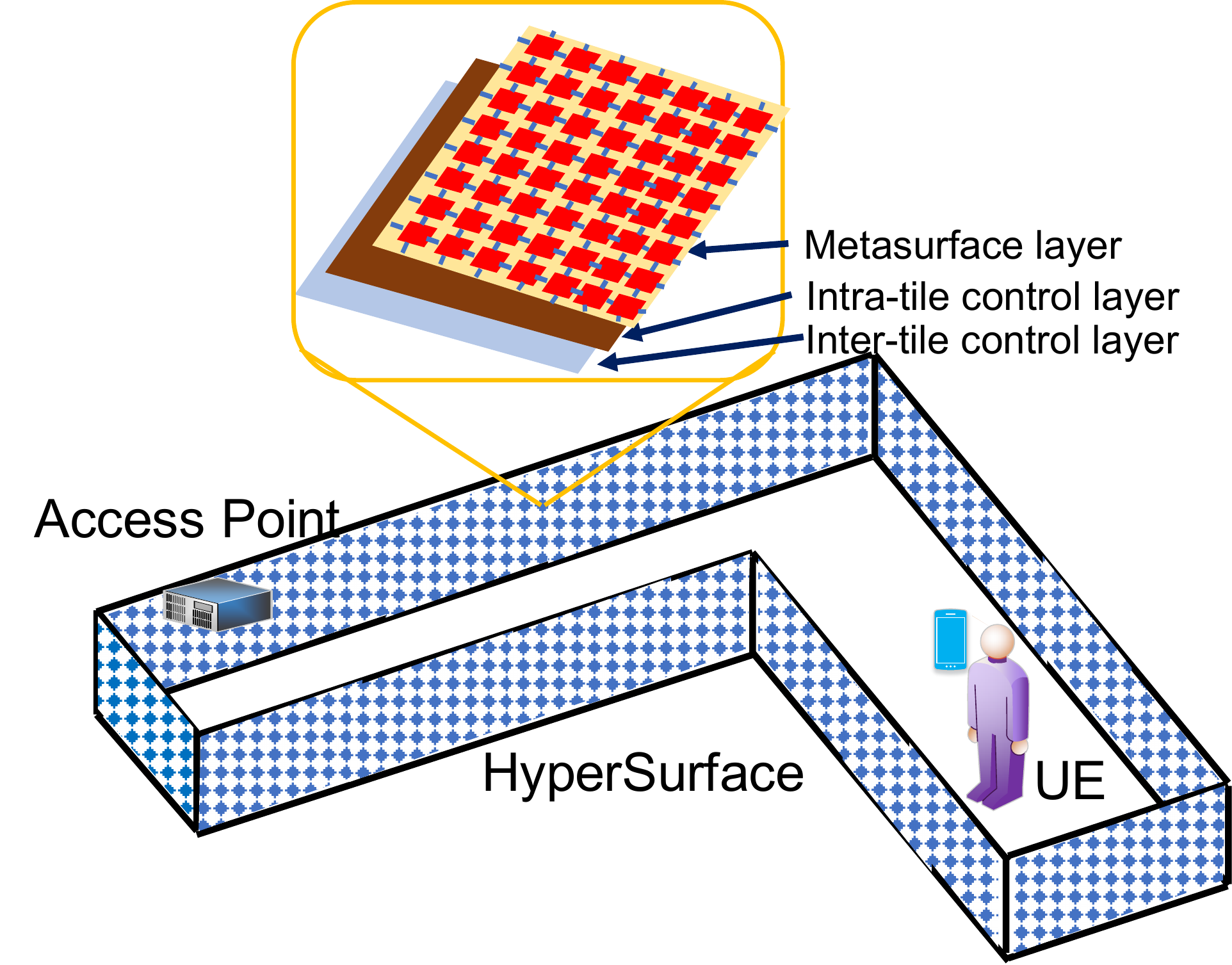} \label{fig:hypersurface}}       
        \subfigure[]{ \includegraphics[trim = 0mm 0mm 00mm 0mm, clip, width=0.38\textwidth]{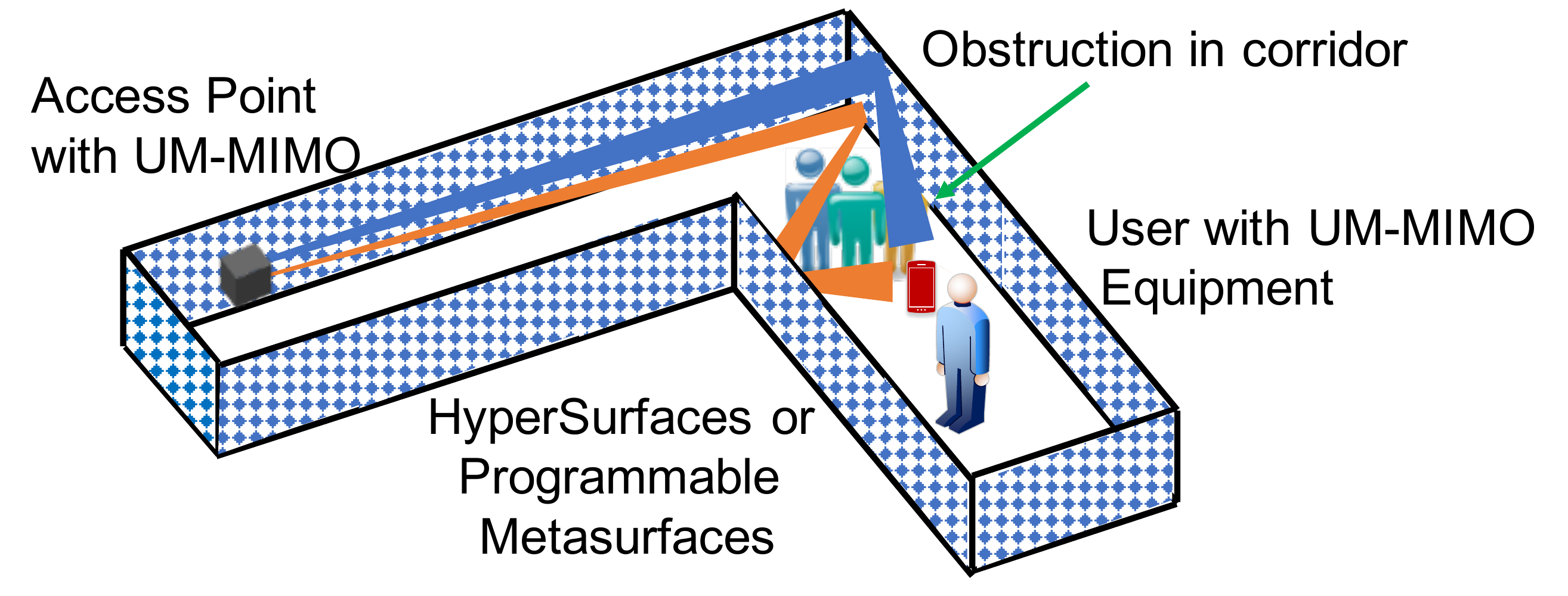} \label{fig:joint}}     
\caption{The illustrations of our designs based on proposed techniques. (a) A use case of graphene-based reflectarrays. The reflectarrays are jointly operated with the base station to reflect propagated signals and boost the strength to desired directions.. (b) The HyperSurfaces as shown in blue rectangular arrays. The enlarged area in a yellow block demonstrates the structure of metamaterial consists of meta-atoms that can be dynamically controlled to tune the HyperSurface tiles to desired angles. Even when the user (shown in purple) has no LOS link to the access point (shown in dark blue), the surrounding HyperSurface tiles will be tuned in 3D angles to direct signals through reflections. (c) The joint design based on UM-MIMO communication, reflectarray or HyperSurface, and the distance-adapative modulation technique.}
        \label{fig:design}
\end{figure*}

In environments with dense multipaths at mm-wave and THz bands, reflectarrays can be deployed to serve single and multiple transmitter and receiver pairs to communicate simultaneously and to extend the transmission range. For example, in an indoor environment where the direct path from a transmitter to a receiver is blocked, as shown in Fig.~\ref{fig:RefArray}, a reflectarray close to the THz source can be used as a reflector to bounce off the signal towards the UE. The reflectarray can dynamically tune the phase of the elements that can sense the transmitted signal to direct the reflected rays towards the users, without any complicated signal processing techniques at the UE side. Additionally, since multiple reflectarray elements will form sharp beams targeting specific users, the interference among users will be mitigated. Compared to the HyperSurface, reflectarrays are an economic choice for improving coverage probability and extending transmission distance for its relative easiness of installation. The reflectarrays can be installed close to access points (APs), or around turning points or blockage areas.

However, reflectarrays also show some limits in their application. First, the efficiency of electronic tuning is highly dependent on the array size and the characteristics of the environment. Especially in mm-wave and THz bands where the signal transmission paths can be easily distorted by any movement in the environment, the time efficiency and accuracy of channel estimation is critical in providing satisfying link quality to users.   Second, at THz band the material for building reflectarrays needs to be reconsidered because studies show that 120 GHz is deemed as the upper limit for micro-electro-mechanical systems (MEMS) which are the most commonly seen in current antenna architecture~\cite{hum2014reconfigurable}.  

%\begin{figure}[!t]
%\centering
%        \includegraphics[width=0.45\textwidth]{reflectarray_new.png}
%        \caption{The illustration of a use case of graphene-based reflectarrays. The reflectarrays are jointly operated with the base station to reflect propagated signals and boost the strength to desired directions.}
%        \label{fig:RefArray} 
%\end{figure}

\section{HyperSurfaces}\label{sec:surface}
Beside the intelligent surfaces which target to control the electromagnetic behavior of the environment or with sub-wavelength resolutions, recent research advances are aimed at controlling the characteristics of the propagation environments in order to improve the transmission distance and solve the non-line-of-sight (NLOS) problem. In this direction, the concept of HyperSurfaces or software-defined meta-surfaces has been recently proposed~\cite{liaskos2015design}. In the near future where IoT connects billions of devices, the reflectarrays and HyerSurfaces will serve as optimal solutions to satisfy the exponential growth in system throughput. Use cases include indoor meeting rooms or corridors with multiple sensors and devices connected and a single AP cannot satisfy the connectivity requirements, as shown in Fig.~\ref{fig:RefArray} and~\ref{fig:hypersurface}.

The HyperSurface is a new type of planar metasurface which can be coated on the surface of indoor environments and can be controlled via software programs to change its EM behavior. The key technology is enabled by metasurfaces, which consist of hundreds of element called meta-atoms, a conductor with the size smaller than half wavelength. Metasurfaces can control the EM waves that impinge on it at certain frequency bands at a very high spatial resolution. These elements are networked by a set of miniaturized controllers that connect the switches of the metasurfaces and a gateway serves as the connectivity unit to provide inter-element and external control~\cite{liaskos2015design}. As illustrated in Fig.~\ref{fig:hypersurface}, the signal propagation routes can be optimized for each communication link with the novel design of HyperSurface tiles using metamaterials. Compared to reflectarrays, metasurfaces can exhibit unconventional electromagnetic properties by interacting with electromagnetic waves at a sub-wavelength scale. Metasurfaces allow one to manipulate incoming waves in ways that are not possible with conventional reflectarrays, including wave steering, wave absorption, and wave polarization~\cite{liaskos2015design}.

%\begin{figure}[!t]
%\centering
%        \includegraphics[width=0.35\textwidth]{hypersurface_new.png} 
%        \caption{An illustration of HyperSurfaces which is shown in blue rectangular arrays. The enlarged area in a yellow block demonstrates the structure of metamaterial consists of meta-atoms that can be dynamically controlled to tune the HyperSurface tiles to desired angles. Even when the user (shown in purple) has no LOS link to the access point (shown in dark blue), the surrounding HyperSurface tiles will be tuned in 3D angles to direct signals through reflections.}
%        \label{fig:hypersurface} 
%\end{figure}

%\subsubsection{Architectures of HyperSurface}
Based on the design of HyperSurface, there are three layers to realize the functionalities of communications, namely, the metasurface layer, the intra-tile control layer, and the tile gateway layer, respectively~\cite{liaskos2015design}. Specifically, the metasurface layer is comprised of the meta-atom with sub-wavelength size whose configuration is adjustable according to the EM function. The meta-atoms are controlled by active CMOS switches and passive conductive patches. The intra-tile control layer is the electronic hardware component that can control the HyperSurface via software. This layer comprises a network of multiple controllers, each of which is connected to an active meta-atom element.  In particular, the key information, such as switch configurations among tiles, is exchanged within the network of controllers. The tile gateway layer determines the communication protocols between the controller network in the intra-tile control layer and the external network. This architecture of three layers guarantees flexibility and accuracy in the operation workflow of HyperSurface.

In general, hundreds of tiles are embedded behind large areas of walls. As in the default mode, all tiles keep parallel to the walls in which they are embedded and also perpendicular to the ground. When activated, the networked tiles will turn to various angles in both azimuth and elevation in order to provide possible reflection paths from the transmitter to the receiver. Through optimization process in the third layer, all tiles will eventually find their best angles in azimuth and elevation to maintain paths, minimize path loss, mitigate undesired multipaths, and extend the transmission distance.

%\subsubsection{Application of HyperSurface}
As opposed to traditional reflectors, HyperSurface serves as a delicate medium to reflect EM waves with ultra-high spatial resolution, and the dominant advantage of HyperSurface is the flexible control brought by the networked architecture of metasurfaces. Additionally, HyperSurface is embedded behind common construction materials which will not affect the aesthetics design and layout of the environment, while reflectarrays are often occupying much space and influence its appearance.  

In an environment of a wireless transmitter/receiver pair operating in a general indoor scenario, there normally exists both NLOS and LOS areas. Compared to the LOS area, the NLOS area experiences lower SNR which leads to worse coverage and lower data rates. Additionally, the multipath fading phenomenon becomes dominant, since only reflected, diffracted, and scattered paths with greatly attenuated power and random phase can be found in such NLOS area. In order to form constructive multipath propagation and mitigate the destructive distorted multipath fading, appropriate reflection functions at each tile are needed in order to yield a focusing lens behavior. Thus, the power received by user equipment (UE) in NLOS areas can be considerably higher than the default behavior, meanwhile the propagation paths become more uniform and extended in distance~\cite{liaskos2015design}.

\section{Joint Design of the Four Directions}\label{sec:joint}
To combine the benefits of the individual technology, we exploit the possibilities to form the joint force and target for maximum distance improvement. In this section, we propose a joint design of the four directions. 
Despite the uniqueness of each technique, the joint design based on their strengths can yield even better performance. Specifically, for example, since HyperSurface and reflectarrays are based on the same foundation of ray optics and they all act as reflectors, they can be dynamically combined to formulate a hybrid intelligent surface with the capabilities to accommodate different user density and resolution requirement. In real-world design and implementation, the intelligent surfaces can be divided into different areas to adapt to specific user density and resolution requirements.  

%\begin{figure}[!t]
%\centering
%        \includegraphics[width=0.5\textwidth]{joint.pdf}
%        \caption{The illustration of a joint design based on UM-MIMO and reflectarrays or HyperSurfaces.}
%        \label{fig:joint} 
%\end{figure}

Moreover, as illustrated in Fig.~\ref{fig:joint}, UM-MIMO can be deployed at both AP and UE to work along with HyperSurface to ease the signal processing overhead and increase the time efficiency. Even when the UE is in motion, the HyperSurface can provide reliable link to the dynamic channel. Additionally, the distance-adaptive design, which provides optimal resource allocation design for multi-band channels, can be compatible with UM-MIMO or HyperSurface to increase the spectral efficiency in mm-wave and THz band networks. For example, with the high beamforming and spatial multiplexing gain brought by UM-MIMO,  the adaptively modulated signal is more robust when propagated in complicated environments in both indoor and outdoor scenarios, thus extending the transmission distance. 

%Last but not the least, since the four directions focus on different aspects of can be coordinated altogether to achieve the optimal performance. Since HyperSurface utilizes novel metamaterial and will be embedded in construction walls while reflectarrays are padded on surface of walls, they can overlap on the same area and operate simultaneously. Meanwhile, the UM-MIMO at the BS will amplify the signal adaptively modulated based on the distance information from UEs' feedback in order to maintain a constant signal level, even under the condition of multiple user case. The systematic design and implementation will consider all factors from the four directions, and the gain from this joint design will surpass that from deployments of individual technique. 
\begin{figure*}[!t]
\centering
	\subfigure[]{ \includegraphics[trim = 0mm 0mm 0mm 0mm, clip, width=3.2in]{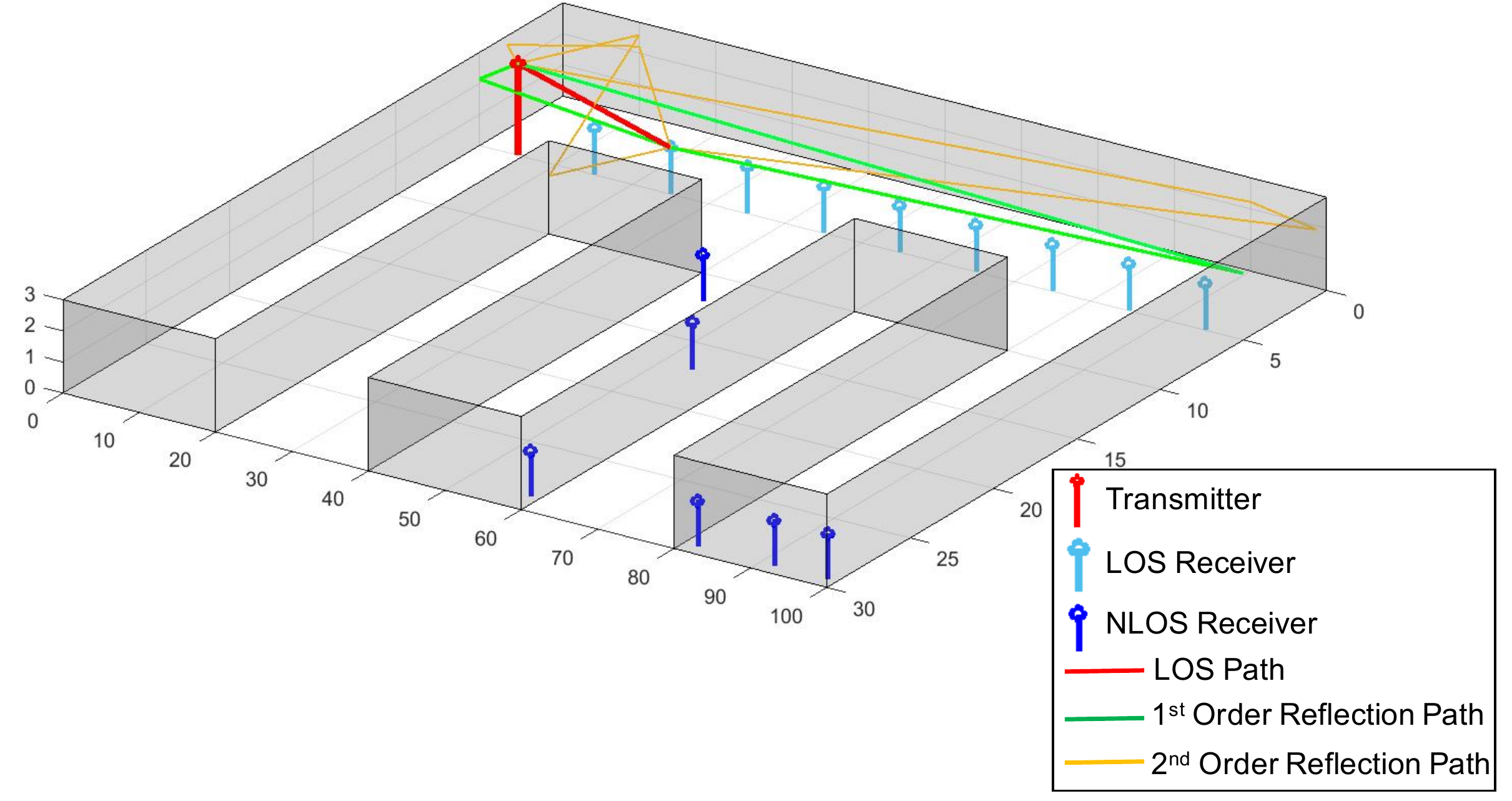} \label{fig:layout}}
%        \subfigure[]{ \includegraphics[trim = 10mm 10mm 10mm 0mm, clip, width=1.6in]{ummimo_10.eps} \label{fig:um_sim}}
%        \subfigure[]{ \includegraphics[trim = 10mm 10mm 10mm 0mm, clip, width=1.6in]{HS_sim_10.eps} \label{fig:HS_sim} }
%        \subfigure[]{ \includegraphics[trim = 10mm 0mm 10mm 0mm, clip, width=2.8in]{joint_14.eps} \label{fig:joint} }
        \subfigure[]{ \includegraphics[trim = 0mm 0mm 0mm 0mm, clip, width=3.2in]{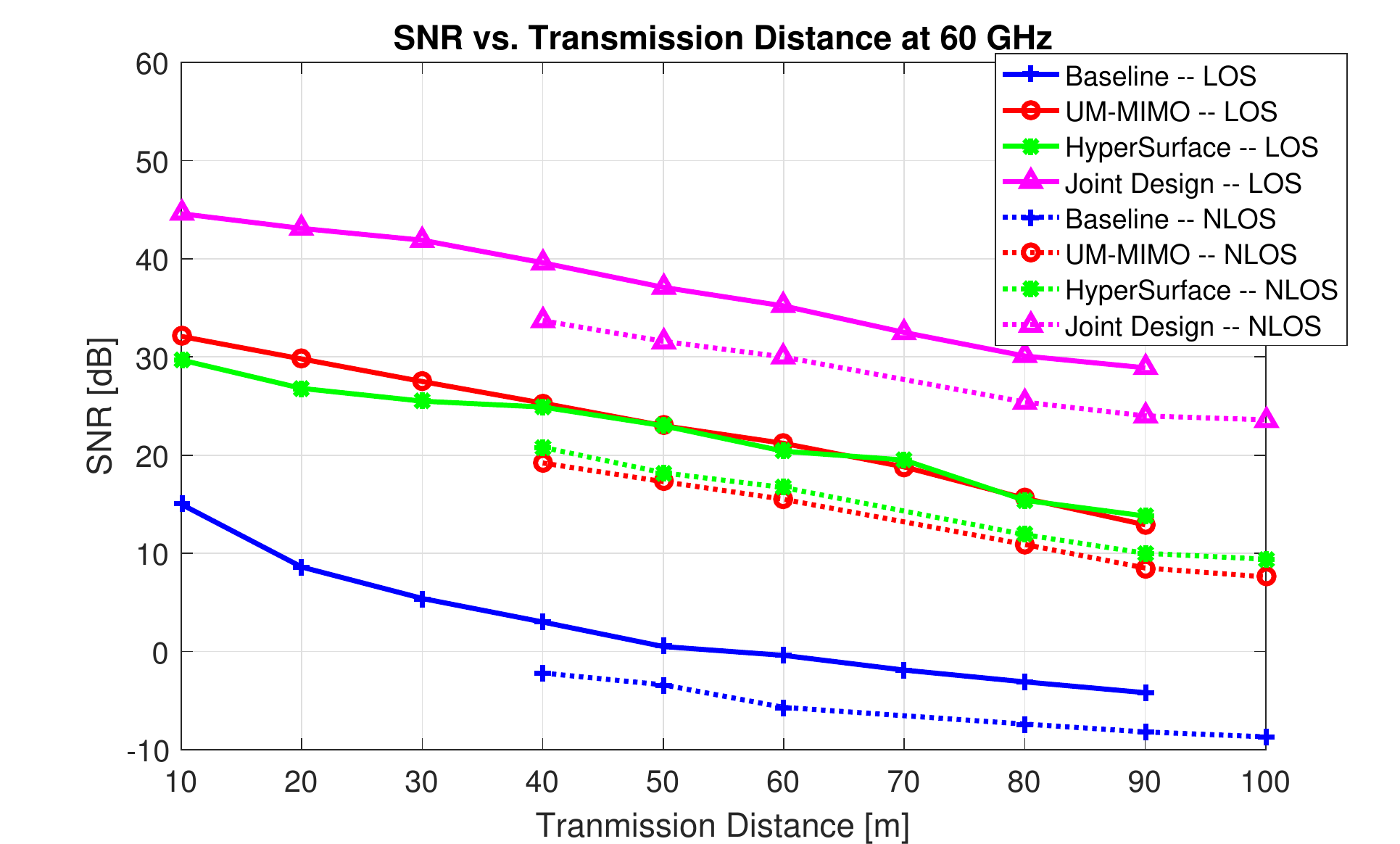} \label{fig:trend60ghz} }       
        \subfigure[]{ \includegraphics[trim = 0mm 0mm 0mm 0mm, clip, width=3.2in]{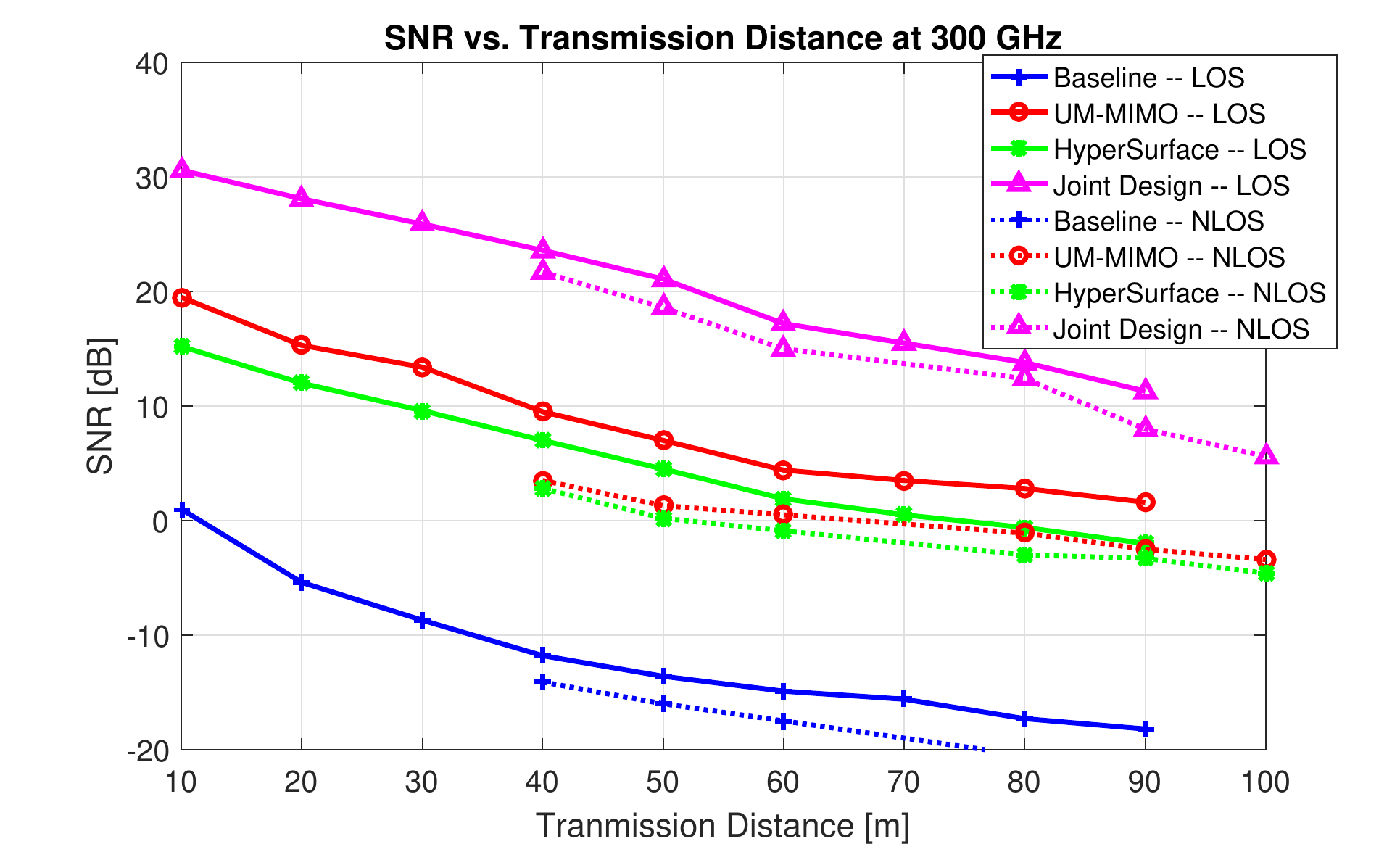} \label{fig:trend300ghz} }   
        \subfigure[]{ \includegraphics[trim = 0mm 0mm 0mm 0mm, clip, width=3.2in]{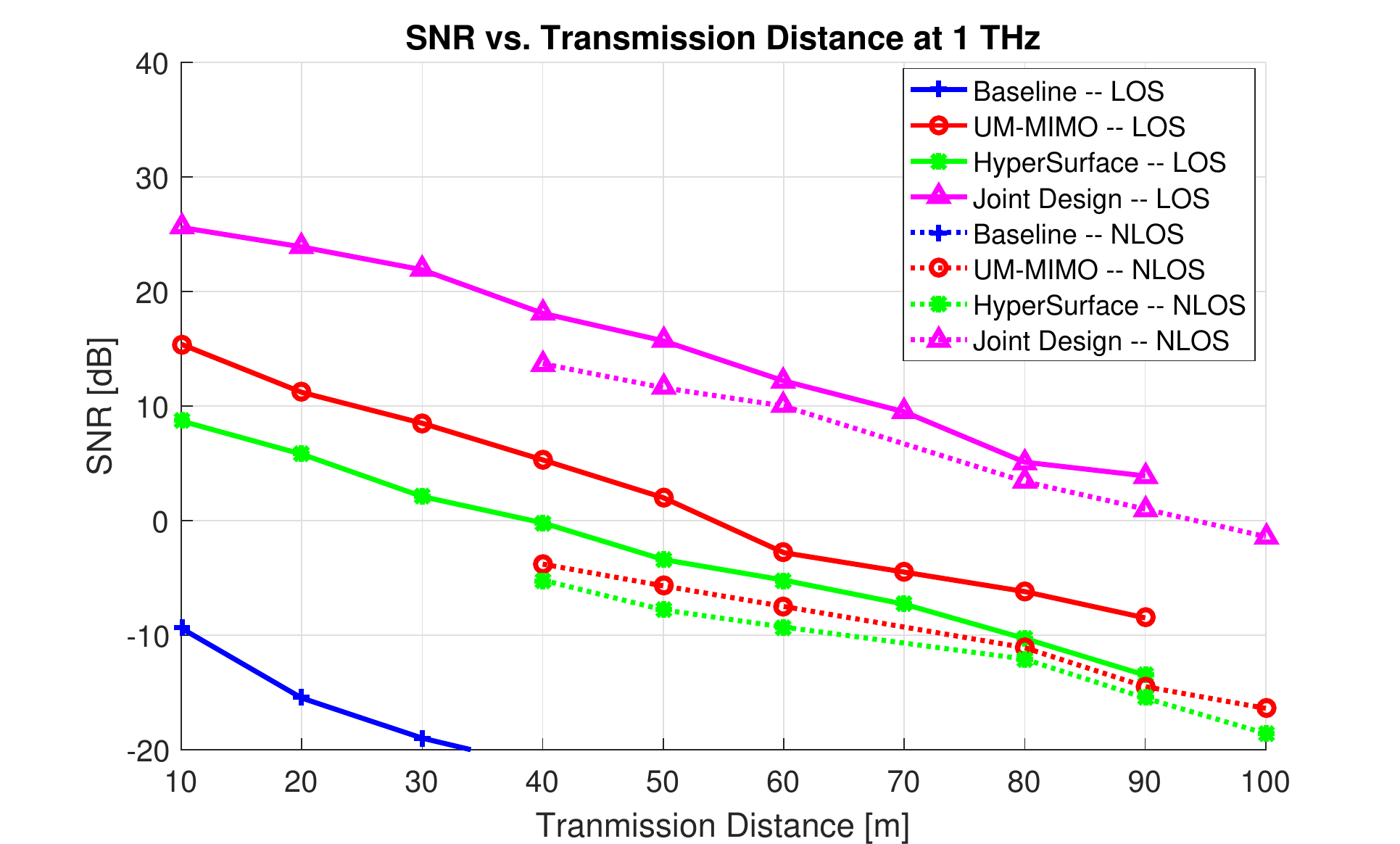} \label{fig:trend1thz} }    
\caption{Simulation results in an indoor hallway with one Tx and 15 Rx locations. (a) The layout and transceiver locations of the ``E'' shape simulation environment with ray-tracing results shown for a pair of Tx-Rx link in the LOS area. The LOS path is shown in red color, and reflected paths are shown in various colors to represent multipaths from first and second order reflections. (b)-(d) Comparison of SNR at various distance with and without utilizing different techniques (UM-MIMO, HyperSurface, and the joint design) at center frequencies of 0.06, 0.3, and 1~THz, respectively.}
        \label{fig:sim}
\end{figure*}

\section{Performance Evaluation}\label{sec:evaluation}
In this section, we provide numerical results to quantitatively illustrate the benefits of the aforementioned four technologies in Sec.~\ref{sec:modulation}~--~\ref{sec:surface} and the joint design in Sec.~\ref{sec:joint} to extend the communication distance. The simulation environment is an ``E'' shape indoor hallway as shown in Fig.~\ref{fig:layout}, which resembles the layout of the Van Leer building in School of ECE at Georgia Tech. In particular, an AP is utilized to provide ultra-broadband THz communications to the UEs along the hallway. When users are blocked by walls, LOS transmissions are obstructed and only paths through reflections can provide viable links. In our simulation environment, the indoor space has a height of $\mathrm{H = 3~m}$. All walls, floors, and ceilings are regarded as planar surfaces comprised of concrete and plywood coverings. 

Specifically, as shown in our simulation environment in Fig.~\ref{fig:layout}, there are in total 15 Rx locations including nine in LOS area and six in NLOS areas, as well as one Tx location located in the hallway on the ceiling level. The Rx locations are chosen in order to compare the same distance in both LOS and NLOS areas. Due to the specific environment layout, some Rx with distances less than 40~m and of 70~m cannot be found in NLOS areas. The nine Rx in LOS area have the distance of 10~m to 90~m from the Tx, while six Rx in NLOS areas have the distance of 40--60~m, and 80--100~m from the Tx, all with a spacing of $\mathrm{10~m}$. The Tx is located at $\mathrm{(5~m, 5~m, 2.95~m)}$ and the Rx has a uniform height of $\mathrm{h = 1.5~m}$.  We utilize a three-dimensional dynamic ray-tracer developed for mm-wave and THz bands, specifically to perform ray tracing from the Tx to all Rx locations in 3D space~\cite{terarays}, which has been used to support the simulation work in~\cite{han2015multi}.  This map-based ray-tracer is built based on 3D channel models in mm-wave and THz frequency bands and is validated against field measurements. With the capability to configure the simulation environment based on real-world scenarios in both indoor and outdoor space, the ray-tracer provides accurate and realistic spatial channel properties with detailed path information on LOS, reflection, diffraction, and scattering paths. In this simulation, we evaluate three distinct frequency bands at 0.06, 0.3, and 1~THz, and the bandwidth is set as $10\%$ of the center frequency in each simulation case. The transmitted signal power is set to be 10~dBm at the Tx. In the case of UM-MIMO, the gain of antenna arrays is 30~dBi at both Tx and Rx where in the baseline case, omnidirectional antennas are used at both sides. Without loss of generality, in our simulation, we assume the reflectance values of the walls, floors, and ceilings are 0.75, 0.5, and 0.8, respectively~\cite{han2015multi}. The noise power spectral density is $-160$~dBm/Hz~\cite{jornet2011channel}.  

%This is due to the large path loss caused by the Tx-Rx separation distance, and also the reflection loss resulted from multiple bounces off the walls. After deploying UM-MIMO on the AP, all Rx locations have good coverage and the transmission distance is extended by 10 times further. With optimization algorithms performed based on the hallway, the wall surfaces are padded with HyperSurface to also extend the transmission distance by 10 times. Since HyperSurface can direct the reflection paths to their desired directions, all Rx locations are covered with a strong improvement. Transmission distance improvement is significant that even further Rx at 100 m from Tx can be covered with decent strength.

Based on the 3D ray-tracing simulation results at 0.06, 0.3, and 1~THz, we are able to examine the transmission distance improvement in mm-wave and THz frequency bands. As shown in Fig.~\ref{fig:trend60ghz}, the distance to reach a threshold of SNR of 10~dB is around 18~m in the baseline case at 60 GHz in LOS (with no techniques utilized). After applying either the UM-MIMO communication technique, HyperSurface, or the joint design, the distance to reach the same threshold of SNR is increased to more than 90~m. Even for Rx located in NLOS areas, the distance improvement is significant. Similar trend can be observed in both cases at 300~GHz and 1~THz. As shown in Fig.~\ref{fig:trend300ghz}, at 300~GHz, the SNR threshold of 10~dB cannot be reached without any techniques deployed, while the joint design provide satisfying link up to 85~m for UEs in both LOS and NLOS areas. Moreover, the joint design offers robust transmission until 60~m for both cases even at 1~THz, as shown in Fig.~\ref{fig:trend1thz}. The simulation results also show that the distance improvement of these technologies and the joint design is consistent throughout mm-wave and THz bands, with the average gains introduced by UM-MIMO communication, HyperSurface, and the joint design to be approximately 17~dB, 15~dB, and 32~dB, respectively.

\section{Conclusion}\label{sec:conclusion}
This paper analyzes the distance limitation problem faced by both mm-wave and THz bands caused by atmospheric attenuation and molecular absorption in wireless communication systems. Through in-depth studies on current research advances on physical layer distance-adaptive design, ultra-massive MIMO, reflectarrays, and HyperSurfaces, we demonstrate directions to solve the limited transmission distance problem. Moreover, joint design based on the four directions is highlighted to fully address the issue of limited transmission range and blockage effect. Based on the analyses, simulation results demonstrate at least five times distance improvement in utilizing individual technology and 10 times improvement in joint design in real-world communication scenarios.

% or



% \section*{Acknowledgment}

\ifCLASSOPTIONcaptionsoff
  \newpage
\fi



%\bibliographystyle{IEEEtran}

\bibliography{ComMag_v10_short_Rev2.bib}

\begin{thebibliography}{10}
\providecommand{\url}[1]{#1}
\csname url@samestyle\endcsname
\providecommand{\newblock}{\relax}
\providecommand{\bibinfo}[2]{#2}
\providecommand{\BIBentrySTDinterwordspacing}{\spaceskip=0pt\relax}
\providecommand{\BIBentryALTinterwordstretchfactor}{4}
\providecommand{\BIBentryALTinterwordspacing}{\spaceskip=\fontdimen2\font plus
\BIBentryALTinterwordstretchfactor\fontdimen3\font minus
  \fontdimen4\font\relax}
\providecommand{\BIBforeignlanguage}[2]{{%
\expandafter\ifx\csname l@#1\endcsname\relax
\typeout{** WARNING: IEEEtran.bst: No hyphenation pattern has been}%
\typeout{** loaded for the language `#1'. Using the pattern for}%
\typeout{** the default language instead.}%
\else
\language=\csname l@#1\endcsname
\fi
#2}}
\providecommand{\BIBdecl}{\relax}
\BIBdecl

\bibitem{akyildiz20165g}
I.~F. Akyildiz, S.~Nie, S.-C. Lin, and M.~Chandrasekaran, ``{5G roadmap: 10 key
  enabling technologies},'' \emph{Computer Networks}, vol. 106, pp. 17--48,
  2016.

\bibitem{ghosh2014millimeter}
A.~Ghosh, T.~A. Thomas, M.~C. Cudak, R.~Ratasuk, P.~Moorut, F.~W. Vook, T.~S.
  Rappaport, G.~R. MacCartney, S.~Sun, and S.~Nie, ``Millimeter-wave enhanced
  local area systems: A high-data-rate approach for future wireless networks,''
  \emph{IEEE Journal on Selected Areas in Communications}, vol.~32, no.~6, pp.
  1152--1163, 2014.

\bibitem{kurner2014towards}
T.~Kurner and S.~Priebe, ``{Towards THz Communications-Status in Research,
  Standardization and Regulation},'' \emph{Journal of Infrared, Millimeter, and
  Terahertz Waves}, vol.~35, no.~1, pp. 53--62, 2014.

\bibitem{akyildizPHYCOM}
I.~F. Akyildiz, J.~M. Jornet, and C.~Han, ``{Terahertz Band: Next Frontier for
  Wireless Communications},'' \emph{Physical Communication (Elsevier) Journal},
  vol.~12, pp. 16–--32, March - June 2014.

\bibitem{jornet2014graphene}
\BIBentryALTinterwordspacing
J.~M. Jornet and I.~F. Akyildiz, ``{Graphene-based plasmonic nano-transceiver
  for Terahertz band communication},'' in \emph{The 8th European Conference on
  Antennas and Propagation (EuCAP 2014)}.\hskip 1em plus 0.5em minus
  0.4em\relax IEEE, April 2014, pp. 492--496, {US Patent: 9,397,758 B2}.
  [Online]. Available:
  \url{http://www.acsu.buffalo.edu/~jmjornet/patents/2016/p1.pdf}
\BIBentrySTDinterwordspacing

\bibitem{graphene}
\BIBentryALTinterwordspacing
------, ``Graphene-based plasmonic nano-antenna for terahertz band
  communication in nanonetworks,'' \emph{IEEE Journal on Selected Areas in
  Communications}, vol.~31, no.~12, pp. 685--694, December 2013, {US Patent:
  9,643,841 B2}. [Online]. Available:
  \url{http://www.acsu.buffalo.edu/~jmjornet/patents/2017/p1.pdf}
\BIBentrySTDinterwordspacing

\bibitem{han2016distance}
C.~Han and I.~F. Akyildiz, ``{Distance-aware bandwidth-adaptive resource
  allocation for wireless systems in the Terahertz band},'' \emph{IEEE
  Transactions on Terahertz Science and Technology}, vol.~6, no.~4, pp.
  541--553, 2016.

\bibitem{jornet2011channel}
J.~M. Jornet and I.~F. Akyildiz, ``{Channel Modeling and Capacity Analysis of
  Electromagnetic Wireless Nanonetworks in the Terahertz Band},'' \emph{IEEE
  Transactions on Wireless Communications}, vol.~10, no.~10, pp. 3211--3221,
  Oct. 2011.

\bibitem{akyildiz2016ummimo}
I.~F. Akyildiz and J.~M. Jornet, ``{Realizing Ultra - Massive MIMO (1024x1024)
  Communication in the (0.06 - 10) Terahertz Band},'' \emph{Nano Communication
  Networks (Elsevier) Journal}, vol.~8, pp. 46--54, 2016.

\bibitem{hum2014reconfigurable}
S.~V. Hum and J.~Perruisseau-Carrier, ``{Reconfigurable reflectarrays and array
  lenses for dynamic antenna beam control: A review},'' \emph{IEEE Transactions
  on Antennas and Propagation}, vol.~62, no.~1, pp. 183--198, 2014.

\bibitem{liaskos2015design}
C.~Liaskos, A.~Tsioliaridou, A.~Pitsillides, I.~F. Akyildiz, N.~V. Kantartzis,
  A.~X. Lalas, X.~Dimitropoulos, S.~Ioannidis, M.~Kafesaki, and C.~M.
  Soukoulis, ``Design and development of software defined metamaterials for
  nanonetworks,'' \emph{IEEE Circuits and Systems Magazine}, vol.~15, no.~4,
  pp. 12--25, Fourthquarter 2015.

\bibitem{Han2016wideband}
C.~Han, A.~O. Bicen, and I.~F. Akyildiz, ``{Multi-Wideband Waveform Design for
  Distance-adaptive Wireless Communications in the Terahertz Band},''
  \emph{IEEE Transactions on Signal Processing}, no.~4, pp. 901--922, 2016.

\bibitem{akyildiz2016ultra}
\BIBentryALTinterwordspacing
I.~F. Akyildiz and J.~M. Jornet, ``{Ultra Massive MIMO Communication in the
  Terahertz Band},'' Nov.~3 2016, {awarded on Sept. 7, 2017}. [Online].
  Available: \url{https://www.google.com/patents/US20160323041}
\BIBentrySTDinterwordspacing

\bibitem{terarays}
\BIBentryALTinterwordspacing
I.~F. Akyildiz and S.~Nie, ``{TeraRays: The 3D Channel Simulation Platform for
  5G and Beyond},'' 2016. [Online]. Available:
  \url{http://http://bwn.ece.gatech.edu/projects/terarays/index.html}
\BIBentrySTDinterwordspacing

\bibitem{han2015multi}
C.~Han, A.~O. Bicen, and I.~F. Akyildiz, ``Multi-ray channel modeling and
  wideband characterization for wireless communications in the terahertz
  band,'' \emph{IEEE Transactions on Wireless Communications}, vol.~14, no.~5,
  pp. 2402--2412, 2015.

\end{thebibliography}
\bibliographystyle{IEEEtran}


%\end{IEEEbiography}
\begin{IEEEbiographynophoto}{Ian F. Akyildiz (F'96)}
is the Ken Byers Chair Professor in Telecommunications with the School of Electrical and Computer Engineering, Georgia Institute of Technology, and the Director of the Broadband Wireless Networking (BWN) Laboratory and the Chair of the Telecommunication Group at Georgia Tech. He has received numerous awards from IEEE and ACM. His current research interests are in nanonetworks, Terahertz band communication networks, 5G wireless systems, and wireless sensor networks. He has more than 96+K citations and his H-index is 108.
\end{IEEEbiographynophoto}

\begin{IEEEbiographynophoto}{Chong Han (M'16)}
received Ph.D. degree in Electrical and Computer Engineering from Georgia Institute of Technology, USA in 2016. Since 2016, he is an Assistant Professor with the University of Michigan-Shanghai Jiao Tong University (UM-SJTU) Joint Institute, Shanghai Jiao Tong University, China. He is also a collaborating researcher with Shenzhen Institute of Terahertz Technology and Innovation, China. His research interests include Terahertz band and millimeter-wave communication networks, and Electromagnetic nanonetworks.
\end{IEEEbiographynophoto}

\begin{IEEEbiographynophoto}{Shuai Nie (S'13)}
received the B.S. degree in Electrical Engineering from Xidian University in 2012, and the M.S. degree in Electrical Engineering from New York University in 2014. Currently, she is working toward the Ph.D. degree in Georgia Institute of Technology under the supervision of Prof. Ian F. Akyildiz. Her research interests include the 5G wireless system and Terahertz band communication networks.
\end{IEEEbiographynophoto}
\balance
%
%% insert where needed to balance the two columns on the last page with
%%\newpage


%\vfill


\end{document}